\documentclass[a4paper,11pt]{article}
\usepackage{jheppub}
\usepackage{color}
\usepackage{graphicx}
\usepackage{amsmath}

\def\lsim {~^{<~}_{\sim~}}

\newcommand{\comm}[1]{{\color[rgb]{0,0,0}{#1}}}
\newcommand{\comme}[1]{{\color[rgb]{0,0,0}{#1}}}
\newcommand{\com}[1]{{\color[rgb]{0,0,0}{#1}}}
\newcommand{\commee}[1]{{\color[rgb]{0, 0, 0}{#1}}}

\title{Phase structure and the gluon propagator of SU(2) gauge-Higgs model in two dimensions}

\author{Shinya~Gongyo$^{1,2}$ and Daniel~Zwanziger$^{2}$}
\affiliation{
$^1$Department of Physics, Kyoto University, Kyoto 606-8502, Japan\\
$^2$Department of Physics, New York University, New York, 10003, USA}

\abstract{%
We study \commee{numerically} the phase structure and the gluon propagator \commee{of the} SU(2) gauge-Higgs \commee{model} in two dimensions. First, we calculate gauge-invariant quantities\commee{, in particular the} static potential from Wilson Loop, \commee{the} W propagator\commee{,} and the plaquette expectation value. 
Our results suggest that a confinement-like region and a Higgs-like region appear even in two dimensions. In the confinement-like region, the static potential rises linearly\commee{,} with string breaking at large distances, while in the Higgs-like region, it is of Yukawa type\commee{, consistent with} a Higgs-type mechanism. The correlation length obtained from the W propagator has a finite maximum between these \commee{regions}. The plaquette expectation value shows a smooth cross-over consistent with  the Fradkin-Shenker-Osterwalder-Seiler theorem. From these results, we suggest that there is no phase transition in two dimensions.
We also calculate \commee{a} gauge-dependent order parameter in Landau gauge. \commee{Unlike gauge invariant quantities, the gauge non-invariant order parameter has a line of discontinuity separating these two regions}.
Finally we calculate the gluon propagtor\commee{. We infer from its} infrared behavior that the gluon propagator would vanish at zero momentum in the infinite-volume limit, consistent with an analytical study.
}

\begin{document}
\maketitle
\flushbottom


\section{Introduction}

The Higgs mechanism, where the gauge boson acquires a mass from spontaneous symmetry breaking plays an important role in nature \cite{Higgs:1964ia}. In a superconductor, the Abelian Higgs mechanism, known as the Meissner effect, appears and the magnetic field is expelled. In the standard model, there is the non-Abelian Higgs mechanism and W and Z bosons become massive.

The relation between color confinement and the Higgs phenomenon has also been studied from the perspective of the confinement-Higgs behavior \cite{Fradkin:1978dv,Osterwalder:1977pc,Lang:1981qg,Caudy:2007sf,Montvay:1984wy,Bonati:2009pf,Maas:2010nc,DellaMorte:2013yca,Ferrari:2013lja}. In the SU(2) gauge-Higgs model, \commee{a} confinement-like region and \commee{a} Higgs-like region \commee{are present} in four dimensions. In the confinement-like region, the static potential between the colored charges rises linearly until string breaking by pair production, while in the Higgs-like region it is of Yukawa type\commee{,} with a massive gauge boson \cite{Montvay:1984wy,Knechtli:2000df,Philipsen:1998de}. While a line of first-order phase transition or rapid cross-over \commee{separates} the confinement-like and Higgs-like \commee{regions}, there is always a path connecting the two \commee{regions}, because gauge-invariant quantities may be continued analytically between the confinement-like region and the Higgs-like region according to the Fradkin-Shenker-Osterwalder-Seiler theorem \cite{Fradkin:1978dv,Osterwalder:1977pc}.
 
\comm{There are also lattice studies \cite{Caudy:2007sf,Greensite:2004ke} of the order parameter defined below\comme{,} Eq. (\ref{psi}). This order parameter breaks the gauge symmetry, and so \commee{is zero} without a gauge fixing\comme{, in accordance with} Elitzur's theorem \cite{Elitzur:1975im}. However \commee{for some values of the coupling,} \comme{the order parameter} may be finite in some gauges such as Landau gauge and Coulmob gauge in which case the ``remnant'' global gauge symmetry is spontaneously broken. \commee{A} transition line or \commee{a} rapid cross-over line \comme{separates the confinement-like region from the Higgs-like region}. The gauge non-invariant order parameter $\Psi$ in Eq. (\ref{psi}) \comme{exhibits} a line of discontinuity between these two regions. However, in accordance with the Fradkin-Shenker-Osterwalder-Seiler theorem, gauge-invariant quantities are continuous \comme{across} part of this transition line. Although the gauge non-invariant order parameter seems to show \comme{a} phase transition \comme{where gauge-invariant quantities are continuous}, the \commee{location of the} line of the transition depends on the gauge fixing condition and thus it is not physical \cite{Caudy:2007sf}.} 

To our knowledge, \comme{such} studies have not been carried out in two dimensions. This case is of interest from \commee{a} theoretical and phenomenological viewpoint. From the Coleman theorem and the Hohenberg-Mermin-Wagner theorem \cite{Coleman:1973ci,Hohenberg:1967zz,Mermin:1966fe}, there are no Nambu-Goldstone bosons and no phase transitions except for the Kosterlitz-Thouless \commee{transition} \cite{Kosterlitz:1973xp}. However, in the Higgs case, the Nambu-Goldstone bosons are ``eaten" by the gauge bosons \commee{so} there are no massless bosons\commee{,} and these theorems are evaded. \commee{From} the perspective of a condensed matter system\commee{,} such as a two-dimensional superconductor\commee{, it is of interest to} clarify whether the Higgs mechanism occurs or not \cite{Minnhagen:1987zz}.

The behavior of the gluon propagator in the Landau gauge is also of interest in two dimensions from the perspective of the Gribov problem. According to accurate lattice studies in pure gauge theory, the behavior of the gluon propagator \commee{is strongly dependent} on the dimensionality \cite{Vandersickel:2012tz,Maas:2011se,Bogolubsky:2009dc,Dudal:2010tf,Oliveira:2012eh,Maas:2007uv,Cucchieri:2007rg}.
Three- and four-dimensional gluon propagators \commee{appear} finite at zero momentum, while the two-dimensional gluon propagator \commee{appears to} vanish. This behavior in two dimensions is consistent with the behavior of the same propagator \commee{of} the local and renormalizable action, called the GZ action, that includes a cut-off at the Gribov horizon in pure gauge theory \cite{Gribov:1977wm,Zwanziger:1989mf,Zwanziger:1991gz}. 
In addition to these lattice studies, there has been an analytical study recently which shows that in two dimensions, for gauge fixing inside the Gribov horizon, the gluon propagator vanishes at zero momentum in the infinite-volume limit whether the action includes other fields such as fermion and scalar fields or not \cite{Zwanziger:2012xg}.

The aim of this paper is to investigate \commee{numerically} the phase structure and the gluon propagator of the SU(2) gauge-Higgs model in the Landau gauge in two dimensions. In Sec.\ref{GI}, we calculate the gauge-invariant quantities\commee{, in particular the} static potential, \commee{the} W-propagator and the plaquette expectation value. \commee{In} Sec.\ref{Op}, we compare the result \commee{for} the gauge-invariant quantities with the order parameter in the Landau gauge which is \commee{gauge non-invariant}. In Sec.\ref{Gp}, we show the result of the two-dimensional gluon propagator in the Landau gauge.

\section{SU(2) gauge-Higgs model}
We use the SU(2) gauge-Higgs model with the fixed length of the Higgs field in two dimensions. The lattice action is given by
\begin{align}
S= \beta\sum_P \left(1-\frac{1}{2}\mathrm{Tr} U_{P}\right)
- \frac{\gamma}{2}\sum _{\mu , x}\mathrm {Tr}\left [\phi ^\dagger (x) U_\mu (x) \phi (x + \hat{\mu}) \right ] \label{FHaction}
\end{align}
with $U_\mu(x) \in SU(2) (\mu =1,2)$ the link-variable for the gauge field, $U_P \in SU(2)$ the plaquette and $\phi (x) \in SU(2)$ the Higgs field of the frozen length. This action, with frozen length of the Higgs fields, is formally obtained from the unfrozen action,
\begin{align}
S_l &= \beta\sum_P \left(1-\frac{1}{2}\mathrm{Tr} U_{P}\right) \nonumber \\
&- \frac{1}{2}\sum _{\mu , x}\left(\varphi ^\dagger (x) U_\mu (x) \varphi (x + \hat{\mu})  + c.c. \right) \nonumber \\
&+ \sum _x \lambda' \left(\varphi ^\dagger (x) \varphi (x) -\gamma \right)^2,
\end{align}
with $\varphi (x) = {}^t(\varphi ^1 (x), \varphi ^2 (x))$ in the SU(2) fundamental representation. When we rescale $\varphi (x)$ by $\varphi(x)=\gamma^{1/2}\tilde{\varphi}(x)$ and take the limit $\lambda ' \rightarrow \infty$ with fixed $\gamma$, the radial degree of freedom for the Higgs field is fixed, $\tilde{\varphi} ^\dagger (x) \tilde{\varphi} (x) =1$. This action formally reduces to Eq.(\ref{FHaction}) with the SU(2) matrix, $\phi ^{ab} (x) = (\epsilon ^{ac}\tilde{\varphi} ^{*c} (x) , \tilde{\varphi} ^b (x))$, although the limit may not be smooth. 

In these models, there always seems to be the Higgs mechanism at the classical level. However if we consider the quantum effects, whether it occurs or not depends on $\beta$ and $\gamma$ \cite{Fradkin:1978dv,Montvay:1984wy,Caudy:2007sf,Maas:2010nc}. 
A similar discussion is given for the Higgs mass in Eq. (\ref{FHaction}):
though the Higgs field with the fixed length has an infinite mass for the radial degree at the classical level, it is not necessarily infinite at the quantum level \cite{Montvay:1984wy}.

In our calculation, we generate $U_\mu(x)$ in unitary gauge, $\phi (x) =1$, with the Monte Carlo simulation of the action 
\begin{align}
S= \beta\sum_P \left(1-\frac{1}{2}\mathrm{Tr} U_{P}\right)
- \frac{\gamma}{2}\sum _{\mu , x}\mathrm {Tr}\left [ U_\mu (x)  \right ] .
\end{align}
We then obtain $U_\mu(x)$ and $\phi(x)$ in any gauge by gauge transformation.

In fact, $U_\mu(x)$ and $\phi (x)$ in the Landau gauge are obtained by the local maximization of 
\begin{align}
R= \sum_{x ,\mu} \mathrm{ReTr}\left[ U_\mu (x)\right] 
\end{align}
under the gauge transformation and identifying the gauge function with $\phi (x)$ as used in Sec.\ref{Op} and \ref{Gp}. This gauge corresponds to the minimal Landau gauge, choosing some Gribov copy in the Gribov region, this choice being in general algorithm dependent. 

\section{gauge invariant quantities}
\label{GI}
To investigate the two-dimensional phase structure of the SU(2) gauge-Higgs model with the fixed length, we calculate three gauge-invariant quantities, static potential, W propagator and the plaquette expectation value.
\com{\subsection{phase structure at $\gamma=0$ and $\beta = \infty$}
Before showing numerical results, we discuss the model at $\gamma=0$ and $\beta = \infty$.

At $\gamma=0$, this model reduces to
\begin{align}
S= \beta\sum_P \left(1-\frac{1}{2}\mathrm{Tr} U_{P}\right),
\label{gamma0}
\end{align}
which is a pure Yang-Mills theory. \comme{This} is always confined and no transition \comme{appears at} any $\beta$ .
At $\beta=\infty$, we find $U_P \rightarrow 1$, corresponding to a pure gauge.
Thus after gauge transformation, the action reduces to
\begin{align}
S= - \frac{\gamma}{2}\sum _{\mu , x}\mathrm {Tr}\left [\phi ^\dagger (x) \phi (x + \hat{\mu}) \right ]. \label{beta=infty}
\end{align}
This action corresponds to the two-dimensional Heisenberg model, where a transition does not appear.

Furthermore, according to the Fradkin-Shenker-Osterwalder-Seiler theorem, at small $\beta$, \comme{every} gauge-invariant \comme{quantity is continuous} \cite{Fradkin:1978dv,Osterwalder:1977pc}\comme{, and so} no transition at small $\beta$ appears for gauge-invariant quantities.
}
\subsection{static potential}
\label{Pot}
In two dimensional pure Yang-Mills theory, the Wilson-loop potential is exactly linear. On the other hand, if the Higgs mechanism occurs, the gauge boson becomes a massive vector boson and the potential behaves like
\begin{align}
V(r) &\sim -\int dp \frac{e^{-ipr}}{p^2+m^2}\sim -e^{-mr},
\end{align}
where $m$ is the mass of vector boson. 

We calculated the static potential between two charged sources from the Wilson loop :
\begin{align}
V(r) &= -\lim_{t\rightarrow\infty}\frac{1}{t}\ln W(r,t),
\end{align}
where $W(r,t)$ is the Wilson loop defined as the product of the link variables along an $r\times t$ rectangular loop. \com{ The lattice calculation was performed at $\beta = 120, 200$ and $\gamma =2, 8$.} The static potential was extracted from the Wilson loop by varying the time direction and finding the convergent point of $-\ln W(r,t) /t$.

\begin{figure}[h]
\begin{center}
\includegraphics[scale=0.7]{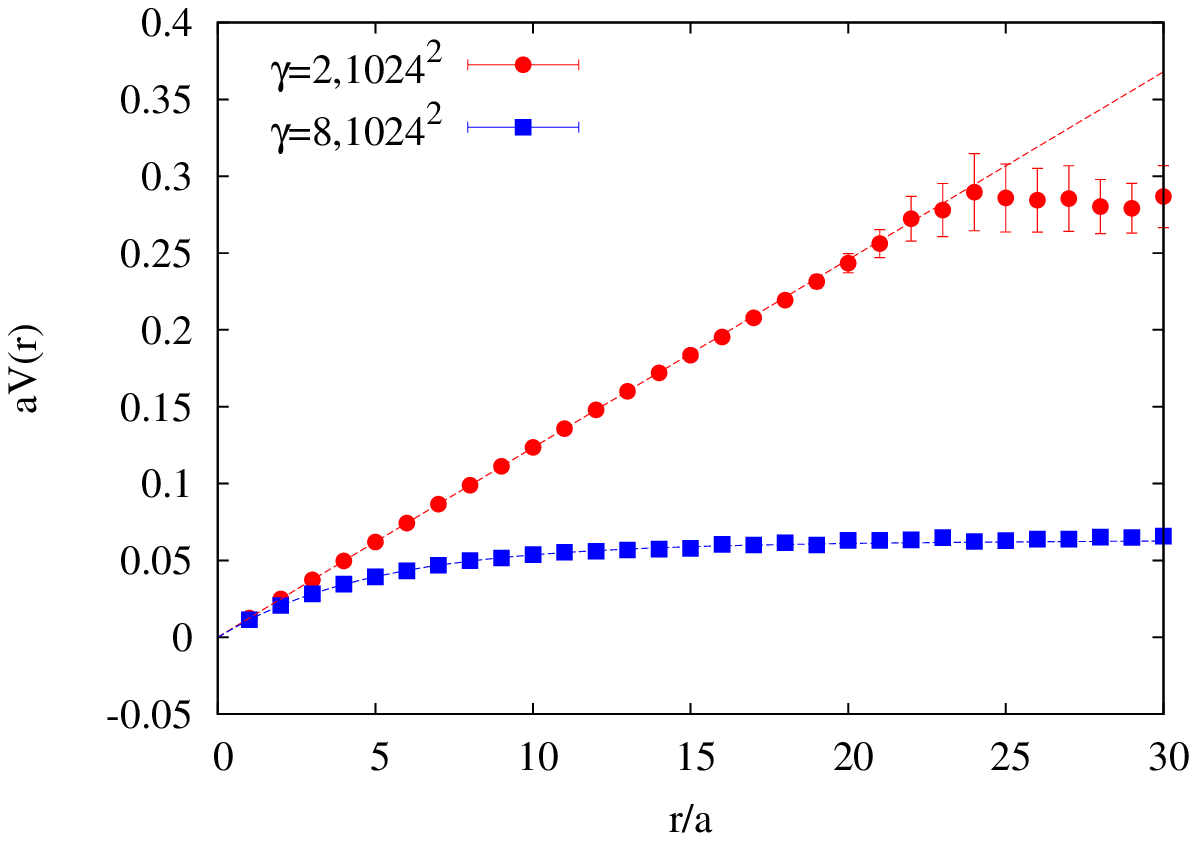}
\includegraphics[scale=0.7]{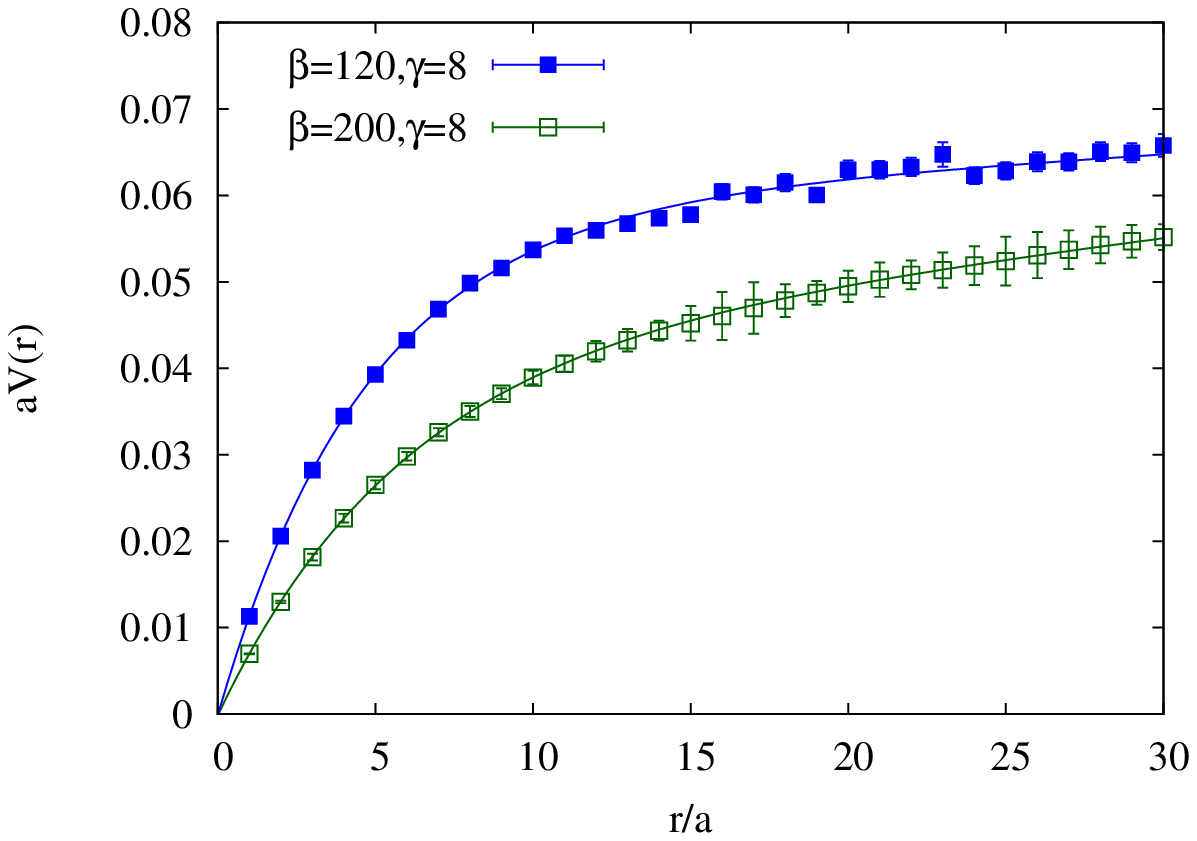}
\caption{
\comm{The static potential at $\beta=120$ and $\gamma = 2,8$ (top), and at $\beta=120,200$ and $\gamma = 8$ (bottom) in lattice units.} The lines are the fit results.}
\label{fig1}
\end{center}
\end{figure}

\com{The static potential at $\beta=120$, and $\gamma = 2$ and $8$ is shown in the top panel of FIG. \ref{fig1}} and the fit results of the form 
\begin{align}
-Ae^{-mr}+ \sigma r + B, \label{fit form}
\end{align}
with the parameter $A,B,m$ and $\sigma$ in the region of $r/a <20$, are summarized in Table \ref{TableI}.

\com{In the top panel of FIG. \ref{fig1}}, the potential at $\gamma =2$ shows the linear potential and the string breaking at $r/a \sim 20$. In contrast, the potential of $\gamma=8$ does not show the linear potential, but behaves like $-e^{-mr}$. Thus $\gamma =2$ corresponds to confinement-like phase and $\gamma =8$ corresponds to Higgs-like phase.

\begin{table}[h]
\caption{ The fit results of the static potential by the fit form Eq. (\ref{fit form}) for $r/a<20$. 
}
\label{TableI}
\begin{center} 
\begin{tabular}{ccccccc}
\hline
\hline
$\beta$ &  $\gamma$     &$ m$ &  $\sigma$ & $A$ & $B$ & $\chi/N$ \\
\hline
120 &$ 2$                &    $0.19(7)$    &  $0.01221(5)$ & $0.0015(7)$ & $0.0015(7)$&$0.3$\\
120 &$8$                 &     $0.207(4)$   &  $7(9)  \times 10^{-5}$ &  $ 0.061(1) $ & $ 0.060 (1) $ &$1.2$\\
200 &$8$                 &     $0.165(2)$   &  $3(4)  \times 10^{-5}$ &  $ 0.044(1) $ & $ 0.044 (1) $ &$0.02$\\
\hline
\hline
\end{tabular}
\end{center} 
\end{table}

\comm{To investigate whether there is a Higgs-like phase, or string breaking at $\gamma=8$, we have taken data also at higher $\beta=200$, corresponding to a finer lattice. Also in this case, our fit shows a very small string tension and is consistent with a potential $V\sim - e^{-mr}$ of a massive particle, as a glance at Table \ref{TableI} shows.}

These behaviors are also supported from the fit results. Though at $\gamma =2$ the mass parameter $m$ is not small, the prefactor $A$ is small and so the first term in Eq. (\ref{fit form}) is negligible compared with the second term $\sigma r$. At $\gamma=8$, the string tension $\sigma$ is almost $0$, and it seems that the potential does not show a linear slope. However because the fit is for $r/a < 20$ and because of string breaking, the difference between the regions may be quantitative (though strong) but not necessarily qualitative.

\subsection{W propagator}
\label{Wp}
We investigated another gauge-invariant quantity, the W propagator, defined by 
\begin{align}
D_{\mu \nu}(x-y) = \frac{1}{3}\sum_{a=1,2,3}\left< W_\mu ^a (x) W_\nu ^a (y)\right>,
\end{align}
where 
\begin{align}
 W_\mu ^a (x) \equiv \frac{1}{i}\mathrm{Tr} \left[\tau ^a \left\{\phi ^\dagger (x)U_\mu (x)\phi (x + \hat{\mu} )\right\}\right] \label{Wfield}
\end{align}
with $\tau ^a$ SU(2) generator. By Eq. (\ref{Wfield}), the W field is gauge invariant and the effective mass for the transverse part of the W field corresponds to the two-dimensional analog of a $1^- $ state \cite{Montvay:1984wy, Maas:2010nc}, which is a singlet state with negative parity. Note that this propagator coincides with the gluon propagator  in the unitary gauge, $\phi (x) =1$.

The effective mass for the W field is estimated from the propagator $D^0_{\mu \nu}(t)$ at zero-spatial-momentum,
\begin{align}
D^0_{\mu \nu}(t) =  \frac{1}{3V}\sum_{a,x_1,y_1}\left< W_\mu ^a (x_1,t) W_\nu ^a (y_1,0)\right>
\end{align}
with the two dimensional volume $V$.
The  effective mass of the transverse part is obtained from $D^0_{11}(t)$ and that of the longitudinal part is from $D^0_{22}(t)$ \cite{Gongyo:2013sha}. In this paper we estimated the transverse effective mass by fitting the linear slope of the logarithm of $D^0_{11}(t)$.
\begin{figure}[h]
\begin{center}
\includegraphics[scale=0.7]{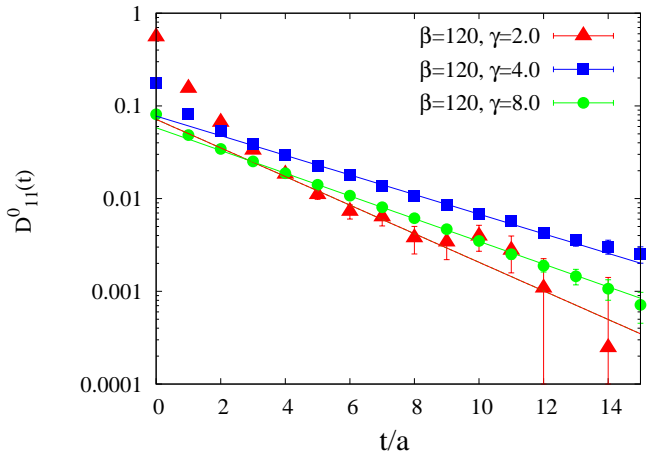}
\caption{\label{fig2}
The zero-spatial-momentum propagator $D^0_{11}(t)$ on $256^2$ at $\beta = 120$ and $\gamma = 2,4$ and $8$ in lattice units.}
\end{center}
\end{figure}

$D^0_{11}(t)$ on $256^2$ lattice at $\beta = 120$ and $\gamma =2,4$ and $8$ are shown in FIG. \ref{fig2}. The logarithm of these zero-spatial-momentum propagators seems to be almost linear or convex upwards, consistent with gauge-invariant fields that satisfy the Kallen-Lehmann representation, in contrast with the gluon propagator in other gauges like the Landau gauge and MA gauge \cite{Mandula:1987rh}. In particular, the two dimensional gluon propagator in the Landau gauge shows notable violation of the Kallen-Lehmann representation as shown in Sec.\ref{Gp} \cite{Cucchieri:2007rg,Maas:2007uv,Zwanziger:2012xg}.
\begin{table}[h]
\caption{ The fit results of the zero-spatial-momentum propagator by $A_D e^{-m_D t}$ for $t/a=4-15$ at $\beta =120$ and on $256^2$ and $512^2$. 
}
\label{TableII}
\begin{center} 
\begin{tabular}{cccccc}
\hline
\hline
$\beta$ &  $\gamma$     &$ m_D$ & $A_D$ & $L^2$& $\chi/N$ \\
\hline
120&$ 2$                &  $0.36(3)$ & $0.07(1)$&$256$&$0.8$\\
120&$4$                 &  $ 0.243(4) $ & $ 0.078 (1) $ &$256$&$0.5$\\
120&$5$                 &  $ 0.240(2) $ & $ 0.069 (1) $ &$256$&$0.1$\\
120&$ 6$                &  $0.265(5)$ & $0.069(2)$&$256$&$0.9$\\
120 &$8$                 &  $ 0.282(1) $ & $ 0.058 (1) $ &$256$&$0.08$\\
120&$10$                 &  $ 0.304(2) $ & $ 0.051 (1) $ &$256$&$0.2$\\
120&$4$                 &  $ 0.243(2) $ & $ 0.077 (1) $ &$512$&$1.0$\\
120&$5$                 &  $ 0.237(3) $ & $ 0.067 (1) $ &$512$&$0.6$\\
120&$ 6$                &  $0.267(2)$ & $0.067(1)$&$512$&$1.0$\\
200&$ 2$                &  $0.52(2)$ & $0.16(1)$&$256$&$1.0$\\
200&$4$                 &  $ 0.207(1) $ & $ 0.065 (1) $ &$256$&$0.2$\\
200&$5$                 &  $ 0.198(1) $ & $ 0.058 (1) $ &$256$&$0.1$\\
200&$ 6$                &  $0.205(2)$ & $0050(1)$&$256$&$0.4$\\
200 &$8$                 &  $ 0.219(1) $ & $ 0.045 (0) $ &$256$&$0.08$\\
\hline
\hline
\end{tabular}
\end{center} 
\end{table}

In Table. \ref{TableII}, we summarize for $D^0_{11}(t)$ the fit analysis of $A_D e^{-m_D t}$ with the parameters $A_D$ and $m_D$ for $t/a=4-15$ \com{at $\beta=120$ and $200$}. The mass parameter $m_D$ corresponding to the physical mass in lattice units decreases in the region of $\gamma<5$, and increases in the region of $\gamma>5$. In other words the mass parameter is minimum at $\gamma \simeq 5$, which means that the correlation is maximum. The volume-dependence of the mass parameters \com{at $ \beta=120$} is also shown in Table. \ref{TableII}. There is almost no volume-dependence between $256^2$ and $512^2$ at $\gamma=4,5$ and $6$ and the minimum does not seem to go to zero. Therefore it looks like a first-order phase transition or cross-over. \com{The fitted W-boson mass values at \comm{$\beta=120$ and $200$ both at $\gamma=8$ in Table. \ref{TableII}} are consistent with the fitted mass values using Yukawa-type potential in Table. \ref{TableI}. \comm{This} indicates that at $\gamma=8$, the potential is given by the massive W-boson exchange, and this region corresponds to the Higgs-like region. This situation is similar to the \comm{four-dimensional} case \cite{Montvay:1984wy}.}

\subsection{Plaquette expectation value}
\label{Pl}
We calculated the plaquette expectation value to see if there is a phase transition or not. The plaquette expectation value is defined as
\begin{align}
\left< P \right> \equiv \frac{1}{2}\left<\mathrm{Tr} U_{P} \right>
\end{align}
with $U_P$ the plaquette. 
This quantity has been investigated in four dimensions to clarify the line of the first-order transition with an end point \cite{Lang:1981qg,Montvay:1984wy,Caudy:2007sf,Bonati:2009pf}. These results are consistent with the Fradkin-Shenker-Osterwalder-Seiler theorem.

\begin{figure}[h]
\begin{center}
\includegraphics[scale=0.6]{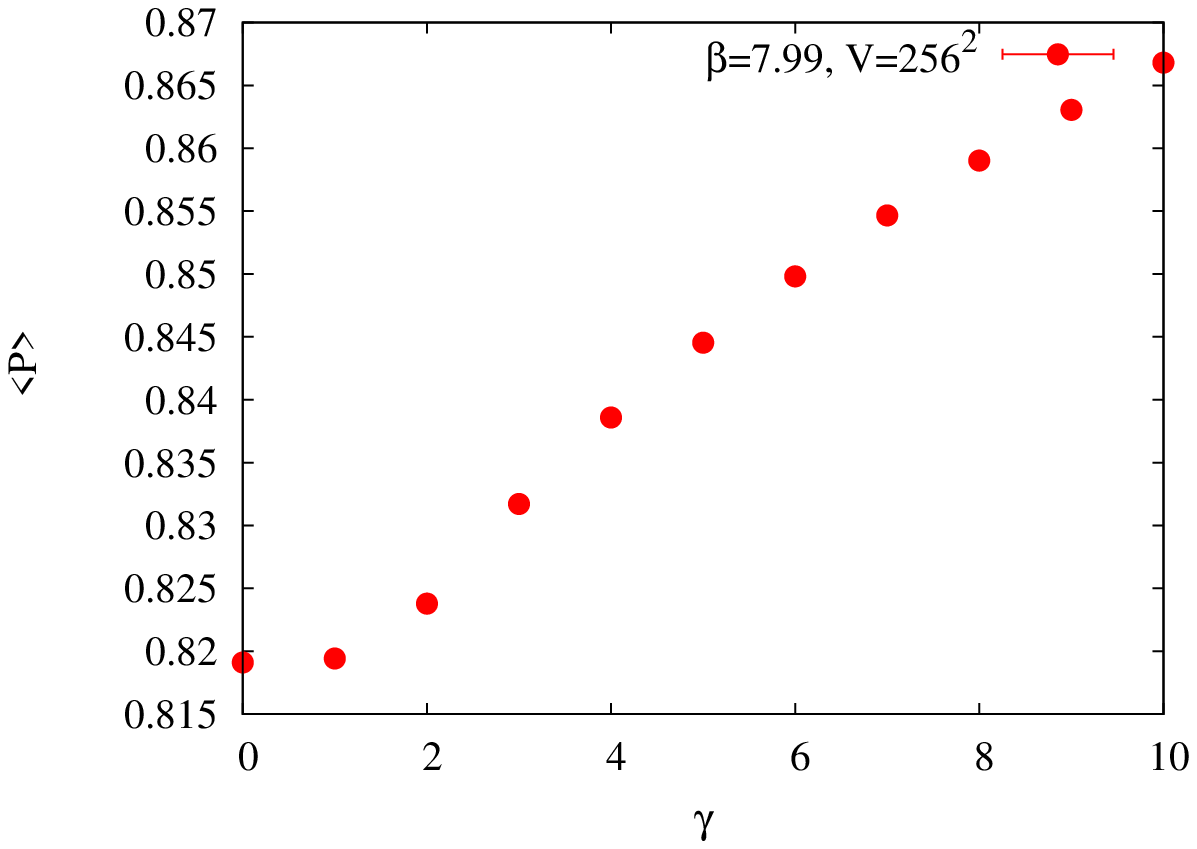}
\includegraphics[scale=0.6]{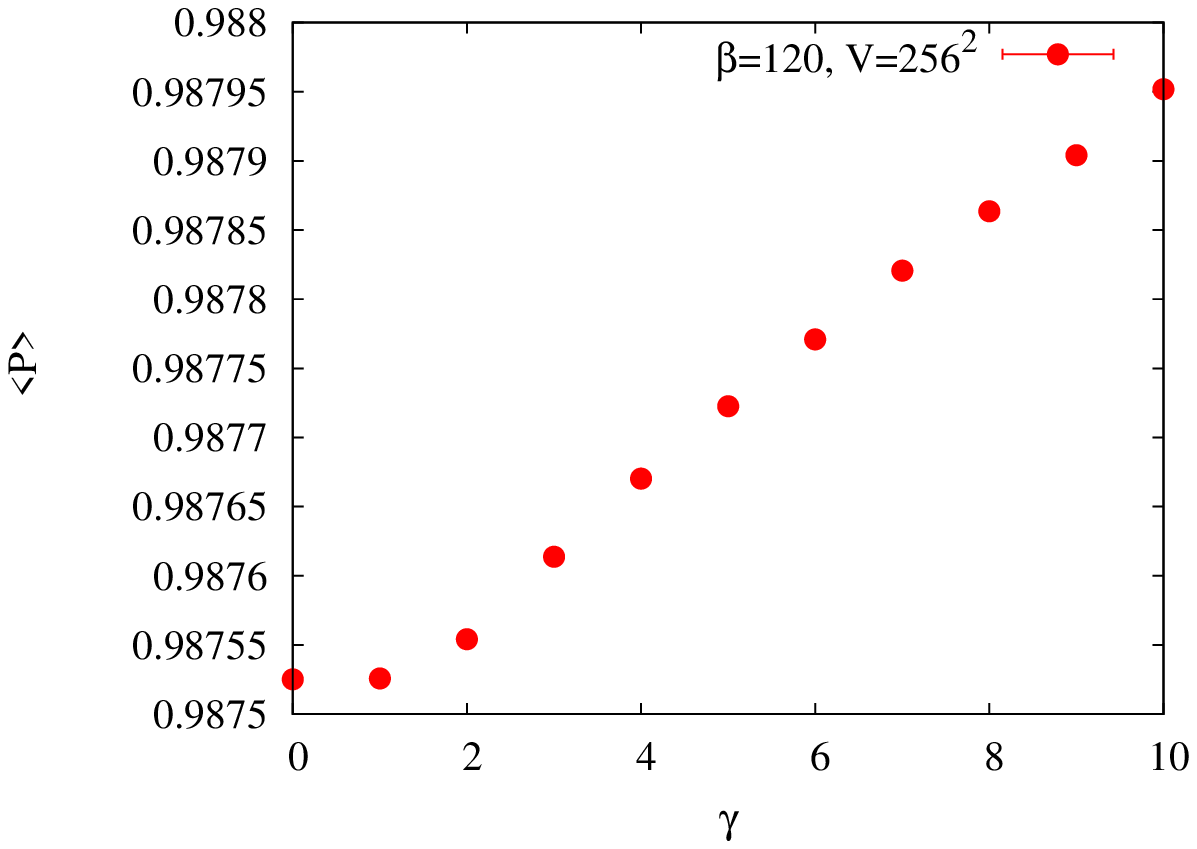}
\caption{\label{fig2.5}
The plaquette expectation value at $\beta = 7.99$ and $120$ on $256^2$}
\end{center}
\end{figure}

 In Fig.\ref{fig2.5}, we show the two-dimensional result at $\beta= 7.99$ and $120$ on $256^2$. The plaquette expectation values at both $\beta=7.99$ and $120$ increase smoothly with increasing $\gamma$ in the region $5 \lsim \gamma \lsim9$ where the gauge-dependent order parameter (that will be discussed next) shows strong variation, and elsewhere. This is indicative of a smooth cross-over (no transition). We also investigated the volume-dependence of some $\gamma$ between $256^2$ and $512^2$ lattice and the difference does not appear. 

Comparing $\beta=7.99$ with $\beta =120$, the behavior of the plaquette expectation value seems to become smooth as $\beta$ increases and thus no transition might appear in the region of $\beta >7.99$.

According to the Fradkin-Shenker-Osterwalder-Seiler theorem, no transition at small $\beta$ appears for gauge-invariant quantities. Therefore also in the region of $\beta < 7.99$, there might be only a smooth cross-over. This accords with the recent work of Cubero and Orland \cite{Cubero:2014hla}, who find that there is no symmetry-breaking Higgs phase in the continuum theory.

\section{Order parameter in the Landau gauge}
\label{Op}
 We calculated the order parameter in the Landau gauge, which does not fix the global SU(2) symmetry.  In four dimensions, in the Higgs-like region, the global remnant gauge symmetry is broken and the order parameter has a finite value \cite{Caudy:2007sf}. 

The order parameter is defined as the magnitude of the spatial average of the Higgs field,
\begin{align}
\Psi  = \frac{1}{2}\mathrm{Tr} \left[ \Phi ^\dagger \Phi \right], \label{psi}
\end{align}
where $\Phi$ is the spatial average of the Higgs field,
\begin{align}
\Phi = \frac{1}{V}\sum _{x} \phi (x).
\end{align}
Note that $\Phi$ is not invariant under SU(2) global-gauge transformations. If the expectation of the magnitude of this quantity, $\left<\Psi\right> $, is zero, the global symmetry is not broken. $\Psi $ is also not invariant under local SU(2) gauge transformations and thus if we do not fix the gauge, then $\left<\Psi\right> =1/V$ by the Elitzur theorem, which approaches zero in the infinite-volume limit \cite{Elitzur:1975im}.  

\begin{figure}[h]
\begin{center}
\includegraphics[scale=0.7]{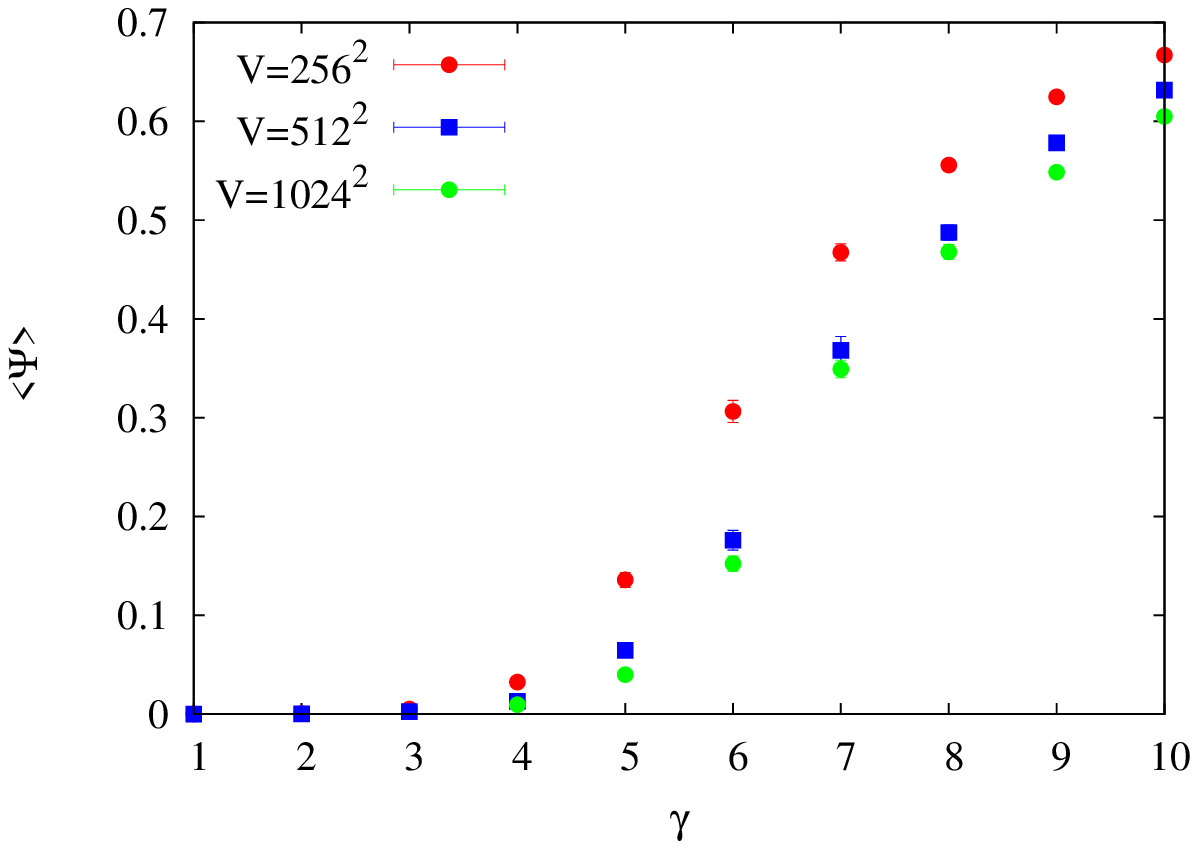}
\caption{\label{fig3}
The order parameter $\left<\Psi \right>=\left<\frac{1}{2}\mathrm{Tr} \left[ \Phi ^\dagger \Phi \right]\right>$ on $256^2,512^2$ and $1024^2$ at $\beta = 7.99$}
\end{center}
\end{figure}

In FIG. \ref{fig3}, we show the order parameter $\left<\Psi\right>$ in the Landau gauge on $256^2,512^2$ and $1024^2$ at $\beta =7.99$. Two regions of small $\gamma$ and large $\gamma$ appear. There appears to be a transition region around $\gamma = 5-8$. However from this data at finite volume we cannot conclude that there is a phase transition to a symmetry-breaking phase at infinite volume, especially because for every $\gamma$, $\left<\Psi\right> $ decreases as the volume increases in contrast to the result of the plaquette expectation value. 

\begin{figure}[h]
\begin{center}
\includegraphics[scale=0.7]{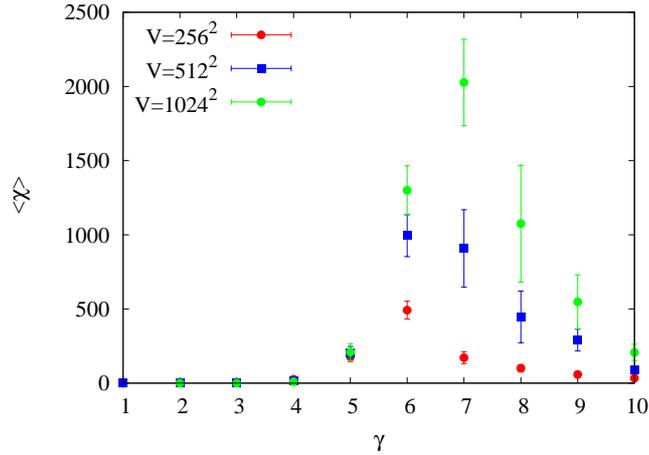}
\caption{\label{fig4}
The susceptibility $\left<\chi\right>$ on $256^2,512^2$ and $1024^2$ at $\beta = 7.99$ in lattice units.}
\end{center}
\end{figure}

To investigate the transition region in more detail, we studied the volume dependence of the susceptibility at $\beta =7.99$,
\begin{align}
\left< \chi \right> \equiv V\left(\left< \Psi ^2 \right> - \left<\Psi \right> ^2 \right).
\end{align}
In FIG. \ref{fig4}, the susceptibility on $256^2, 512^2$ and $1024^2$ lattice is shown. As the volume increases, the height of the peak increases and its width becomes narrow, while the peak point also moves.

Here we consider two possibilities about the order parameter and the susceptibility in the infinite-volume limit: (i) The order parameter approaches a finite value at large $\gamma$ and the peak point of the susceptibility also approaches a finite value or
(ii) the order parameter approaches zero for every $\gamma$ and the peak point of the susceptibility goes to infinity.
In the case of (i), a first-order phase transition or rapid cross-over appears and the Higgs-like region is characterized by the broken phase of the remnant global symmetry in the Landau gauge. This indicates that the Nambu-Goldstone bosons appear, which are not physical particles because of the gauge dependence, and the Coleman theorem cannot be applied to the remnant global symmetry breaking. On the other hand, in the case of (ii), the behavior may indicate Kosterlitz-Thouless type transition.
It is likely that in the Higgs-like region, the Higgs correlator decreases polynomially (instead of exponentially) and this will produce a finite value of $\left<\Psi\right> $ on a lattice of finite volume, which however vanishes in the infinite-volume limit,
\begin{equation}
\langle \Psi \rangle = \lim_{V \to \infty} (2V)^{-1} \sum_x \langle{\rm Tr}[\phi^\dag(x) \phi(0)] \rangle = 0,
\end{equation}
due to the polynomial fall-off. 

Though in order to determine the nature of the transition, further numerical calculations are needed, both cases (i) and (ii) seem to be different from the result of gauge invariant quantities in Sec. \ref{GI}.

\section{Gluon propagator in the Landau gauge}
\label{Gp}
Finally, we investigate the gluon propagator in the Landau gauge. In two dimensional pure gauge theory, the lattice studies show that the gluon propagator in the Landau gauge approaches zero when the momentum goes to zero, which is consistent with the behavior of the GZ action \cite{Cucchieri:2007rg,Maas:2007uv}. Furthermore, the analytical study shows that whether the action includes other fields in addition to gluon fields or not, the two-dimensional gluon propagator must vanish at zero momentum in the infinite-volume limit of lattice gauge theory at fixed lattice spacing for gauge fixing inside the Gribov region only \cite{Zwanziger:2012xg}. 

In this section, we calculate the gluon propagator in the system with Higgs fields given by Eq.(\ref{FHaction}). The momentum-space gluon propagator is defined as
\begin{align}
\tilde{G}_{\mu \nu}(p) = \frac{1}{3}\sum_{a=1,2,3}\left< \tilde{A}_\mu ^a (\tilde{p}) \tilde{A}_\nu ^a (-\tilde{p})\right>,
\end{align}
where $\tilde{p}$ and $p$ are defined as
\begin{align}
\tilde{p}_\mu\equiv \frac{2\pi n_\mu}{aL_\mu},~ p_\mu \equiv \frac{2}{a}\sin \left( \frac{\tilde{p}_\mu a}{2} \right) ,
\end{align}
with $n_\mu = 0,1,2, \ldots ,L_\mu -1$, 
and $\tilde{A}^a_\mu (\tilde{p})$ is defined as
 \begin{align}
\tilde{A}^a_\mu (\tilde{p})  = \sum _x e^{-i\tilde{p}_\nu x_\nu -\frac{i}{2}\tilde{p}_\mu} A^a_\mu (x).
\end{align}
with
\begin{align}
 A_\mu ^a (x) \equiv \frac{1}{i}\mathrm{Tr} \left[\tau ^a U_\mu (x)\right]. \label{Afield}
\end{align}
With this definition, the momentum-space gluon propagator in the Landau gauge has just the transverse part $\tilde{G}(p^2)$ :
\begin{align}
\tilde{G}_{\mu\nu}(p) = \left(\delta _{\mu\nu} - \frac{p_\mu p_\nu}{p^2}\right) \tilde{G}(p^2).
\end{align} 

\begin{figure}[h]
\begin{center}
\includegraphics[scale=0.6]{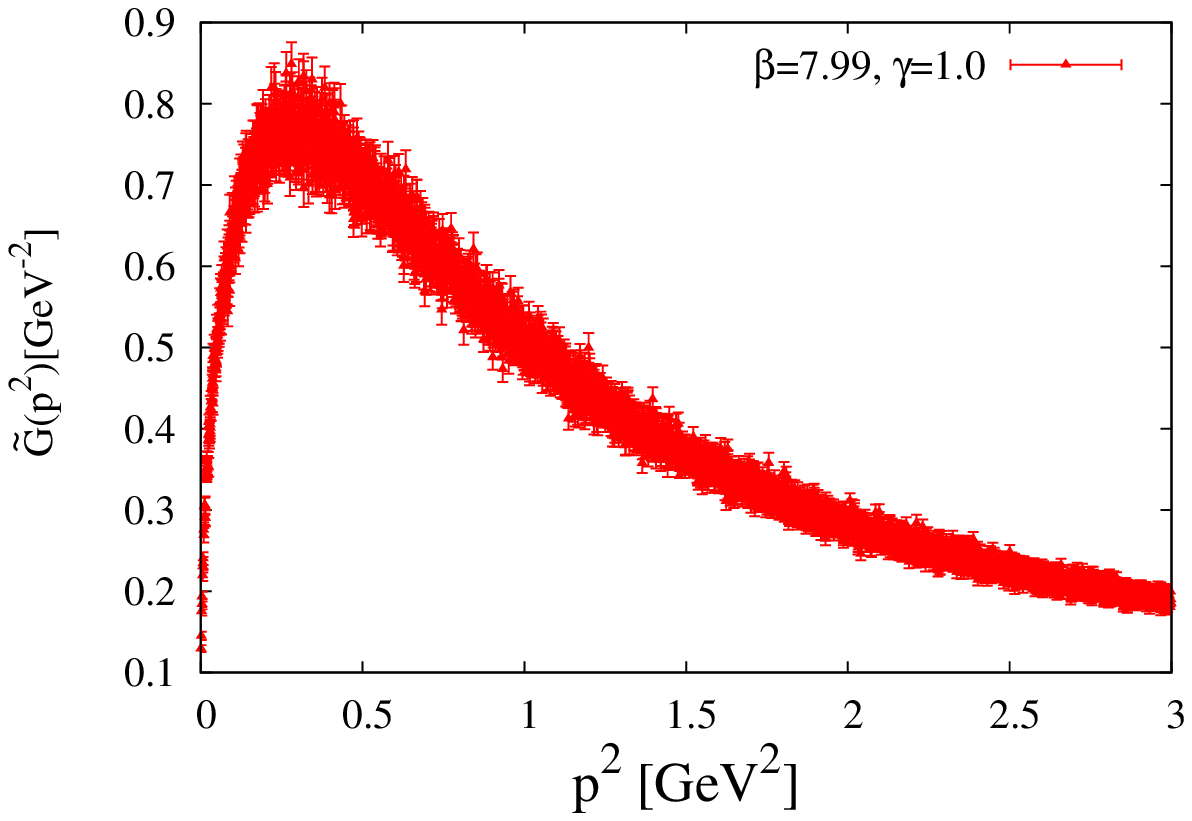}
\includegraphics[scale=0.6]{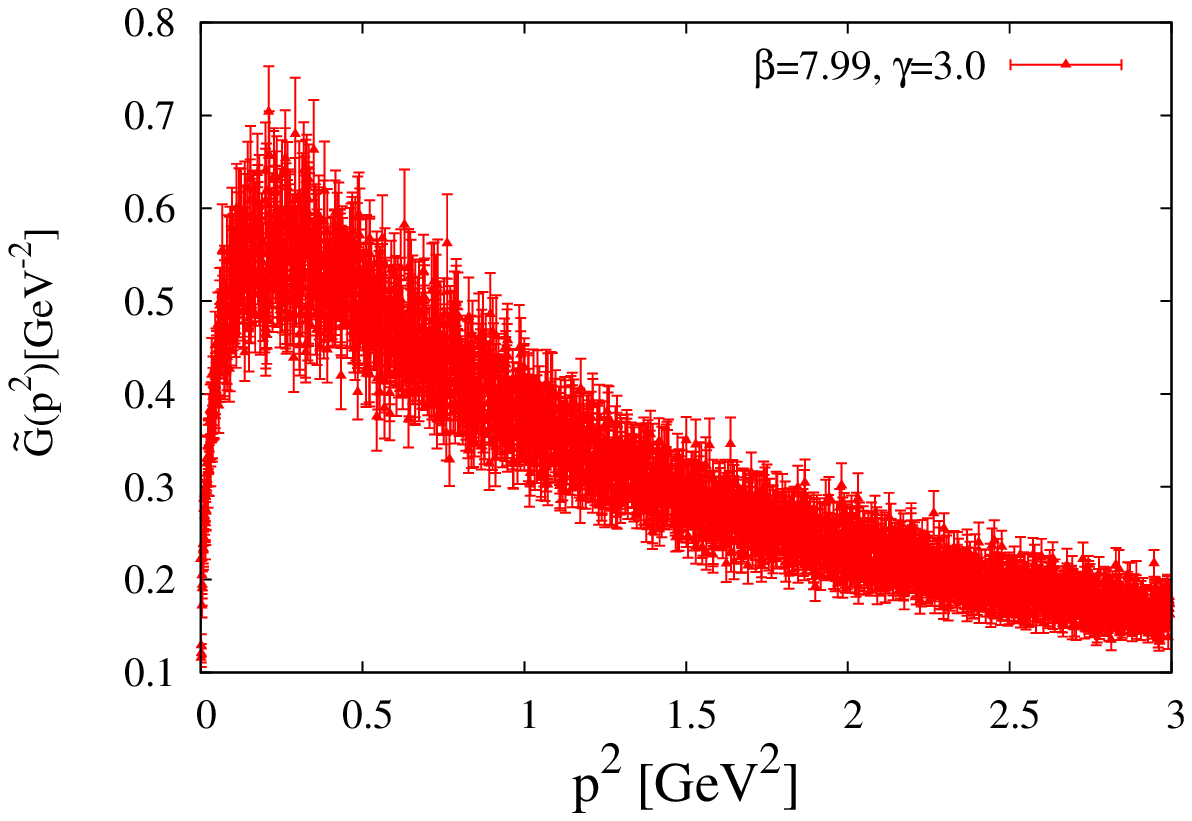}
\includegraphics[scale=0.6]{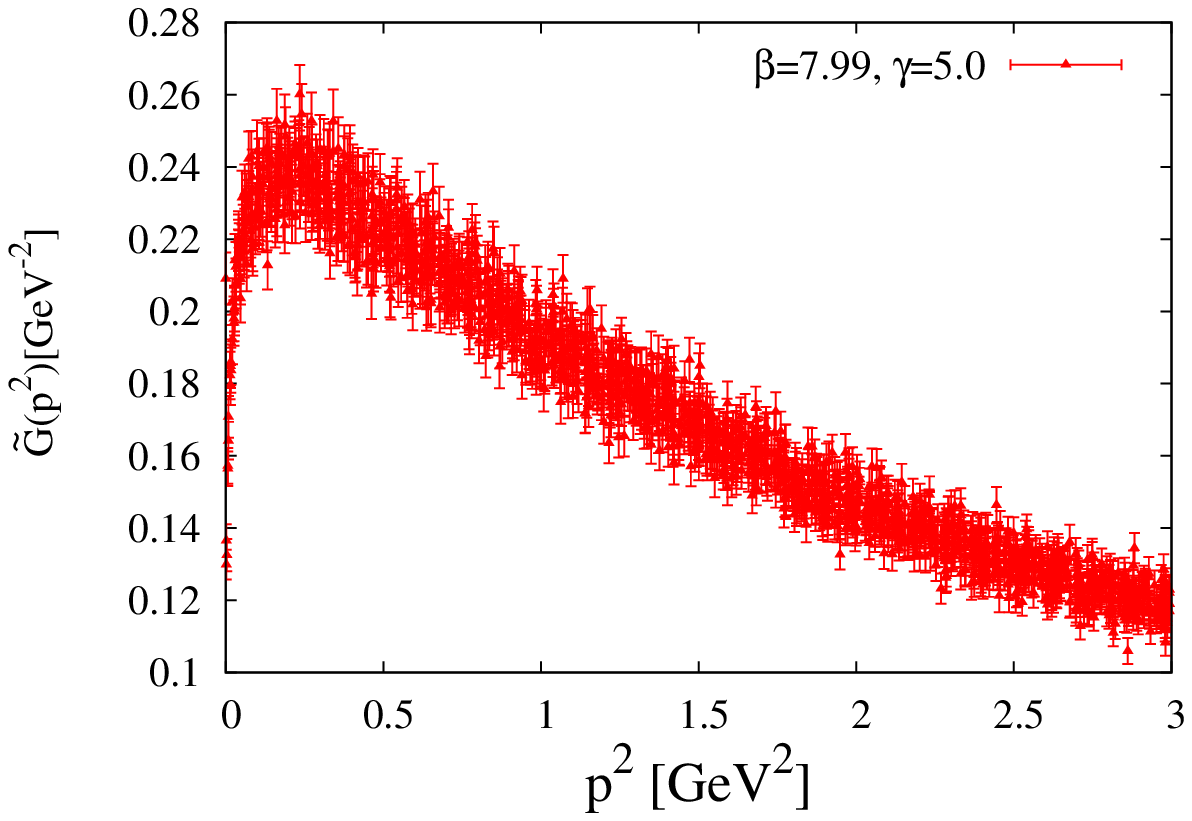}
\caption{\label{fig5}
The gluon propagator in the Landau gauge at $\beta = 7.99$ and $\gamma =1,3,5$ on $256^2$.}
\end{center}
\end{figure}

In Fig.\ref{fig5}, we show the momentum-space gluon propagator in the Landau gauge at $\beta=7.99$ and $\gamma =1,3,5$ on $256^2$. All of the gluon propagators at $\gamma=1,3$ and $5$ decrease rapidly with decreasing momentum, which indicates that they go to zero at zero momentum in the infinite-volume limit. Note that the behavior of these gluon propagators at $\gamma =1,3$ and $5$ seems to be similar qualitatively. 

\comm{We determine the physical scale \commee{so} that the peak point of the propagator in the Landau gage is $0.5$ GeV as shown in Table \ref{TableIII}. In two dimensions, \commee{the} scaling region is determined by $\beta a^2$= const. In fact, the lattice spacing $a$ for the two values of $\beta$ at $\gamma=1$ is consistent with \commee{this} scaling. Furthermore, we need one more physical quantity because of \commee{the other} bare coupling $\gamma$\commee{,} like quark mass in QCD. For example, we can choose $m_D/\hat{p}_{max}=$const. as the physical quantity.}

\begin{table}[h]
\caption{ The peak of the gluon propagator in lattice units, $ \hat{p}_{max}$, and the physical scale. 
}
\label{TableIII}
\begin{center} 
\begin{tabular}{cccc}
\hline
\hline
$\beta  $&$\gamma$     &$ \hat{p}_{max}$ & $1/a[\mathrm{GeV}]$ \\
\hline
$7.99$ &$ 1$                &  $0.46$ & $1.09$\\
$7.99$ &$3$                 &  $ 0.38 $ & $ 1.32 $ \\
$7.99$ &$5$                 &  $ 0.28 $ & $ 1.79 $ \\
$120$ &$ 1$                &  $0.12$ & $4.17$\\
$120$ &$2$                 &  $ 0.11 $ & $ 4.55 $ \\
$120$ &$4$                 &  $ 0.09 $ & $ 5.56 $ \\
\hline
\hline
\end{tabular}
\end{center} 
\end{table}

\begin{figure}[h]
\begin{center}
\includegraphics[scale=0.6]{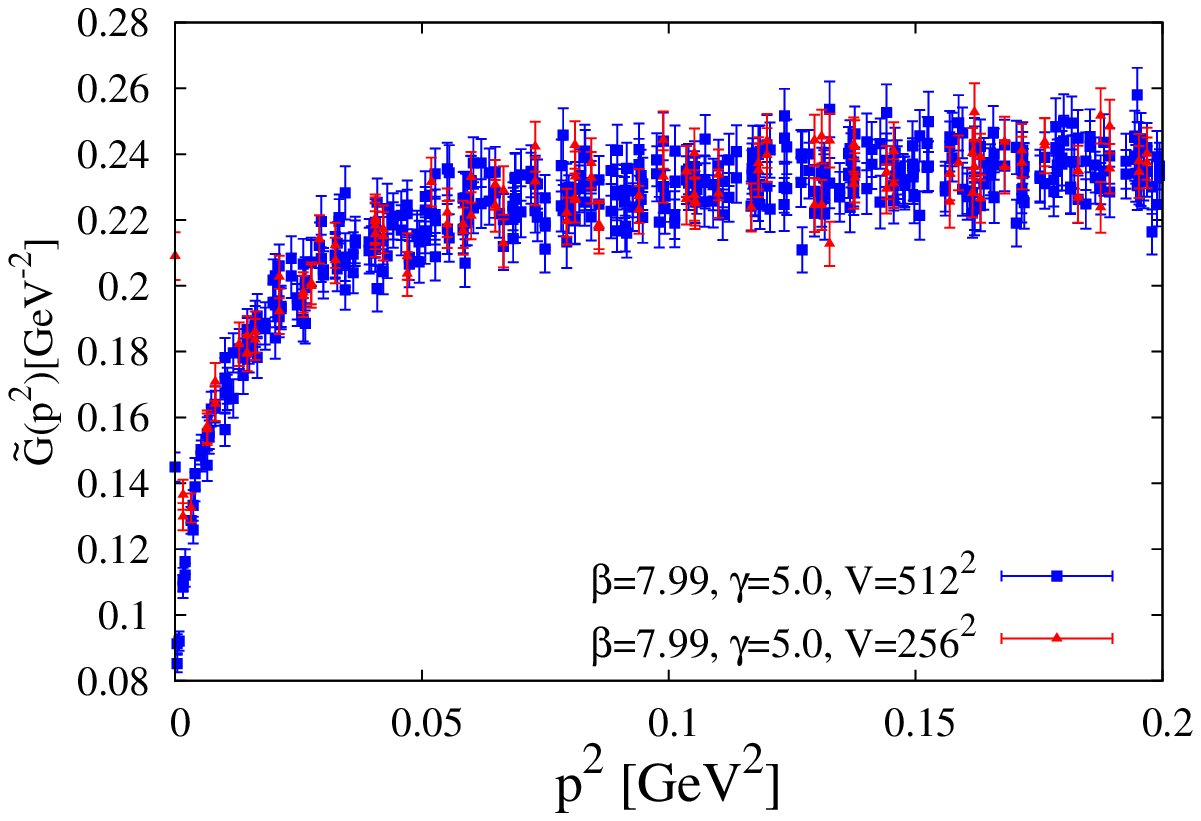}
\includegraphics[scale=0.6]{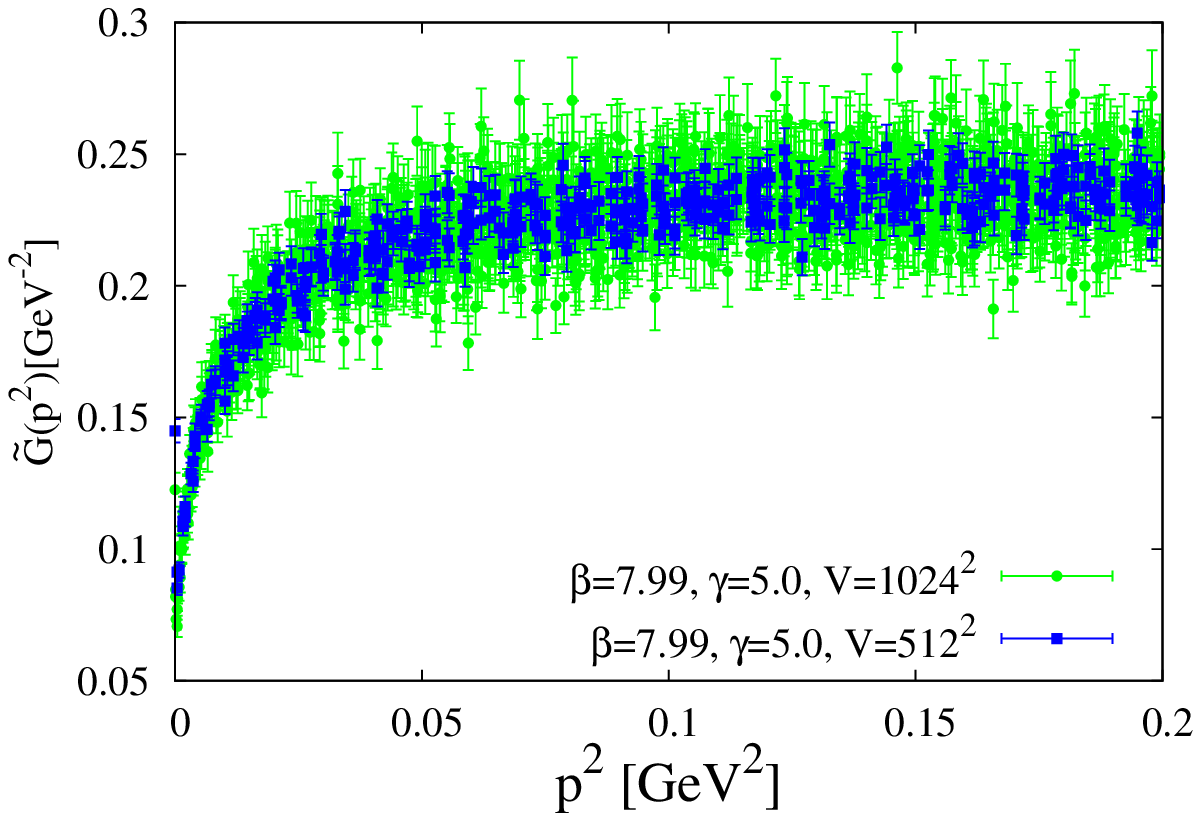}
\vspace*{0.5cm} 
\end{center}
\caption{\label{fig6}
The volume-dependence of the gluon propagator in the Landau gauge at $\beta = 7.99$ and $\gamma =5$ on $256^2, 512^2$ and $1024^2$. These volumes correspond to $(28.6\mathrm{fm})^2, (57.2\mathrm{fm})^2$ and $(114.4\mathrm{fm})^2$, respectively.}

\end{figure}

\begin{figure}[h]
\begin{center}
\includegraphics[scale=0.6]{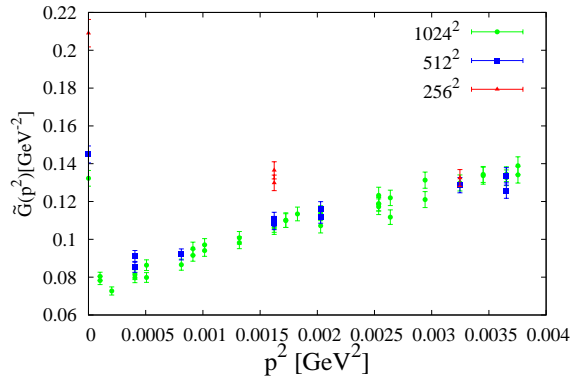}
\end{center}
\caption{\label{fig7}
The infrared behavior of the gluon propagator in the Landau gauge at $\beta = 7.99$ and $\gamma =5$ on $256^2, 512^2$ and $1024^2$.
}

\end{figure}

To elucidate the infrared behavior of the gluon propagator, we investigate the volume-dependence at $\beta=7.99$ and $\gamma = 5$. In Fig. \ref{fig6}, we show the gluon propagators on $256^2, 512^2$ and $1024^2$ in the region of $p^2<0.2\mathrm{(GeV)^2}$. These physical volumes correspond to $(28.6\mathrm{fm})^2, (57.2\mathrm{fm})^2$ and $(114.4\mathrm{fm})^2$, respectively. In Fig. \ref{fig7}, the propagator seems to approach zero with increasing physical volume.

\section{Summary and Concluding Remarks}
First, we studied the phase structure in \commee{the two dimensional} SU(2) gauge-Higgs model.
We carried out the Monte Carlo simulation and calculated gauge-invariant quantities: \commee{the} static potential and the W propagator at $\beta = 120$ \commee{and $200$,} and the plaquette expectation value at $\beta=7.99$ and $120$\commee{, both at various $\gamma$.}

The static potential shows a linear rise and string breaking at $\gamma=2,$ \commee{but looks like a} Yukawa potential at $\gamma =8$. The effective mass of the W propagator in lattice units has a minimum at $\gamma \simeq 5$ which does not go to zero with increasing volume. These results suggest that there is a Higgs-type mechanism \commee{even in two dimensions, with} a confinement-like region \commee{for $\gamma <5$,} and a Higgs-like region \commee{for $\gamma > 5$}. The plaquette expectation value shows a smooth cross-over between \commee{the} confinement-like region and \commee{the} Higgs-like region at $\beta=7.99$ and $120$\commee{,} which is consistent with the Fradkin-Shenker-\commee{Osterwalder}-Seiler theorem. There does not appear to be a phase transition in two dimensions for any gauge-invariant quantities.  We also calculated the \commee{gauge-dependent} order parameter \commee{in the Landau gauge} at $\beta = 7.99$. \comm{As discussed in the Introduction, the} behavior of the \comm{gauge non-invariant} order parameter  seems to show a transition, which \commee{does not happen for} gauge-invariant quantities.

Finally, we have studied the two-dimensional gluon propagator in the Landau gauge at $\beta =7.99$ and $120$ on $256^2$, $512^2$ and $1024^2$. \commee{The} qualitative \commee{behavior is similar for the} various $\gamma$\commee{. They all appear to go to} zero at zero momentum in the infinite-volume limit, consistent with the analytical result \cite{Zwanziger:2012xg}.


\acknowledgments 
The authors are grateful to A.~Maas for suggesting this investigation. They thank J.~Greensite, P.~Hohenberg, A.~Maas, M.~Schaden and A.~Sokal for discussions and comments. 
S.~G.~is supported by the Grant-in-Aid for Scientific Research from the JSPS Fellows (No. 24-1458).   
The lattice calculations were done on NEC SX-9 at Osaka University.




\end{document}